\documentclass[]{spie}  

\usepackage{textcase} 
\usepackage{amsmath,amsfonts,amssymb}
\usepackage{graphicx}
\usepackage[colorlinks=true, allcolors=blue]{hyperref}

\title{Preserving Dense Features for Ki67 Nuclei Detection\footnote{This article is published in SPIE Medical Imaging, 2022, https://doi.org/10.1117/12.2611212.}}

\author[1]{\small Seyed Hossein Mirjahanmardi
\thanks{Corresponding author (email:shmirjahanmardi@ryerson.ca)}}
\author[2]{\small Melanie Dawe}
\author[2]{\small Anthony Fyles}
\author[2]{\small Wei Shi}
\author[2,3]{\small Fei-Fei Liu}
\author[2,4]{\small Susan Done}
\author[1]{\small April Khademi
\thanks{Corresponding author (email: akhademi@ryerson.ca)}}

\affil[1]{\footnotesize Department of Electrical and Computer Engineering, Ryerson University, Toronto, ON, CAN}
\affil[2]{\footnotesize Princess Margaret Cancer Centre, University Health Network, Toronto, ON, CAN}
\affil[3]{\footnotesize Department of Radiation Oncology, University of Toronto, Toronto, ON, CAN}
\affil[4]{\footnotesize Department of Laboratory Medicine and Pathobiology, University of Toronto, Toronto, ON, CAN}

\usepackage{geometry}

\makeatletter
\newcommand{\ps@titlepage}{\let\@mkboth\@gobbletwo
	\def\@oddhead{\vbox to\z@{\kern-\topmargin\kern-.5in
			\hbox{\begin{tabular}{l}
					SPIE Medical Imaging: Digital and Computational Pathology~~~~~~~~~~~~~~~~~~~~Published:04/2022\\
					Vol.~\textbf{12039} (2022), pp.~220-227.
				\end{tabular}\hfil}\vss}}%
	\let\@evenhead\@oddhead%
	\def\@oddfoot{}%
	\let\@evenfoot\@oddfoot}
\def\@AMSclass{}
\def\@keywords{}

\let\ori@maketitle\maketitle
\def\maketitle{\ori@maketitle\thispagestyle{titlepage}}

\let\ori@endabstract\endabstract
\def\endabstract{\par\bigskip%
	\ifx\@AMSclass\empty\else
	\textit{AMS classification:} \@AMSclass.\par
	\fi
	\ifx\@keywords\empty\else
	\textit{Keywords and phrases:} \@keywords.\par
	\fi
	\ori@endabstract}
\makeatother

\begin{document}

\maketitle

\begin{abstract}
Nuclei detection is a key task in Ki67 proliferation index estimation in breast cancer images. Deep learning algorithms have shown strong potential in nuclei detection tasks. However, they face challenges when applied to pathology images with dense medium and overlapping nuclei since fine details are often diluted or completely lost by early maxpooling layers. This paper introduces an optimized UV-Net architecture, specifically developed to recover nuclear details with high-resolution through feature preservation for Ki67 proliferation index computation. UV-Net achieves an average F1-score of 0.83 on held-out test patch data, while other architectures obtain 0.74-0.79. On tissue microarrays (unseen) test data obtained from multiple centers, UV-Net's accuracy exceeds other architectures by a wide margin, including 9-42\% on Ontario Veterinary College, 7-35\% on Protein Atlas and 0.3-3\% on University Health Network. 


\end{abstract}


\section{INTRODUCTION}
\label{sec:intro}  

Breast cancer is the second most commonly diagnosed cancer among women worldwide~\cite{bray} and histopathology has played a pivotal role in its diagnosis, prognostication, and treatment~\cite{shostak}. Traditionally pathologists evaluate excised hematoxylin and eosin (H\&E) stained tissues under microscopes to analyze tissue microstructure, spatial nuclei configuration, and cellular morphology. Grading criteria such as the Modified Bloom-Richardson are used to evaluate nuclear grades and mitotic index \cite{elston}. Recently, the MIKB (Ki67) protein has been gaining momentum as a clinical biomarker to characterize the proliferation and aggressiveness of cancer tumours, which can be used to assist with diagnosis and for treatment planning \cite{dowsett}. This biomarker can potentially improve patient care, but its manual calculation is laborious and time-consuming. The proliferation index (PI) is used to assess the proportion of Ki67 positive cells to the total number of cells in a region. The advent of digital pathology wholeslide scanners has brought the potential to automate Ki67 PI scoring to improve objectivity and turn-around-times (TATs). Deep learning methods have been showing promises in automating this process \cite{srinidhi, van, amgad, graham}, by improving the sensitivity of nuclei detection and increasing robustness to artifacts. For example, piNet \cite{geread_pinet} and KiNet \cite{xing} are two deep learning methods recently developed specifically for Ki67 quantification. While piNet is tested on multi-center datasets, its performance decreases when applied to images with faint nuclei or  close, clustered and overlapping cells. KiNet, developed on pancreatic datasets, is composed of multiple residual connections and multiple stage-level concatenations with F1-score equals 0.804. The reported results were not applied to breast tissues nor tested on a multi-center dataset to assess the generalization of this network.

\indent This paper introduces an enhanced architecture, referred to as UV-Net, that focuses on preserving high-resolution details in pathology images to assess Ki67 PI. The proposed architecture preserves nuclear features through dense "V" blocks to retain the high-resolution details. Experiments are conducted on five multi-center datasets, comprised of tissue microarrays (TMAs) and patches. The results of the proposed system are compared to other conventional architectures. As will be shown, the proposed UV-NET maintains the fine nuclei details which results in significantly higher and consistent performance on multi-center datasets, especially in regions with densely clustered or faint nuclei. Such an approach can significantly contribute to single-cell analysis by scaling the network.

\section{Datasets and Processing Pipeline}
\subsection{Datasets}
The Ki67 stained breast cancer images used in this work are obtained from multiple institutions. The dataset used for training contains 500 patches of size 256$\times$256 extracted from whole slide images (WSI) of two sources: St. Michael's Hospital (SMH) in Toronto, and an open-source "Deepslides" \cite{Senaras} with $\times$20 Aperio AT Turbo and $\times$40 Aperio ScanScope scanners, respectively. The images were annotated with centroid markers that distinguished Ki67 positive and Ki67 negative tumour nuclei cells. This dataset is randomly split into 62$\%$, 20$\%$, and 18$\%$ for training, validation, and testing purposes. To enhance robustness, the partitions contain 12$\%$ nonideal patches including artifacts, non-tumour cells, dust, and overstained regions. Additionally, the two training dataset chosen are in sharp contrast with each other to allow for higher variability of the data. Such a selection plays a key role in generalization robustness.\\
\indent Three other held out datasets are used to test the architectures' generalization capabilities including 411 Tissue Microarrays (TMA)s belonging to 175 patients from University Health Network (UHN) with size 2000$\times$2000, 82 TMAs from Protein Atlas \cite{uhlen} with size 3500$\times$3500, and 30 TMAs from the Ontario Veterinary College (OVC) at the University of Guelph with size 1400$\times$1400. While the OVC data is obtained from canine mammary cancerous tissue, other datasets are all from human breast cancer tissues. The ground truth (GT) for SMH, Deepslides, and OVC data provide individual nuclei cells, while for UHN the proliferation index (PI) is given for each patient (containing 1-3 TMA(s) per patient) annotated by our pathology team. For the Protein Atlas's only PI ranges were available. All images are cropped into tiles with a size of 256$\times$256 RGB, resulting in a total of 48,504 patches for the generalization test.

The ground truth masks include single-pixel nuclear annotations which alleviate the cumbersome process of exhaustive ground truth delineations. However, a single-pixel marker does not show an effective learning performance by the models. We, therefore, symmetrically expand the single-pixel marker to a circular region using a Gaussian function, where the centers of created regions hold the maximum value. Such an approach increases the incorporation of more contextual information from within the nuclei, enhancing the training efficiency. The predicted patches of an image are recombined and the full image is reconstructed before calculating the final PI. This approach reduces the errors due to double counting nuclei on borders. 
\subsection{Processing Pipeline and Frameworks}

\begin{figure}[t]
\begin{center}
   \includegraphics[width=0.6\linewidth]{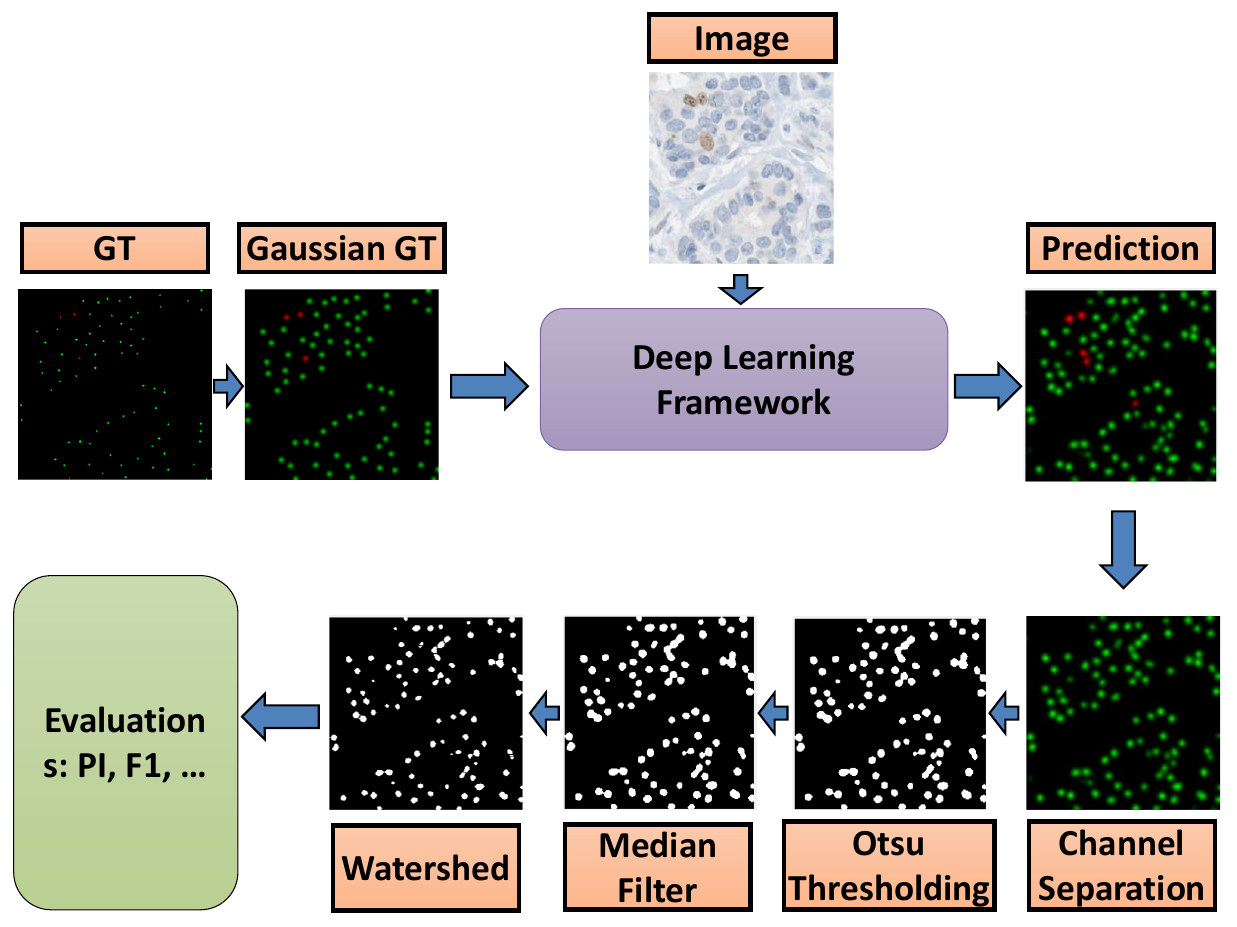}
   \caption{Processing pipeline. The process includes three steps, pre-processing (GT, Gaussian GT), deep learning framework, and post-processing (channel separation, Otsu thresholding, median filter, and watershed).}
\label{fig-Pipeline}
\end{center}
\end{figure}

The entire processing pipeline including the pre and post-processing is shown in Figure~\ref{fig-Pipeline}. The images are patched to size 256$\times$256 and the corresponding Gaussian GT images were created. The SMH and DeepSlides patches were used to train the model, which is detailed in the next section. The predicted images are separated into two channels, one containing Ki67$^{-}$ (green channel), and the other containing Ki67$^{+}$ (red channel). The images are then post-processed. First, the Otsu's thresholding is applied to each channel to convert the regressed prediction into a  binary representation. To remove small and irrelevant false positives, median filtering was applied. The predicted outputs can contain overlapping nuclear regions, therefore, the watershed algorithm is applied to disconnect the overlapped regions. The entire post-processing (Otsu thresholding, median filter, and watershed) is applied to each channel separately. For visualization purposes, Figure~\ref{fig-Pipeline} only focuses on the green (Ki67$^{-}$) channel. The PI is automatically computed as the number of Ki67$^{+}$ divided by the total number of tumour cells as follows:
\begin{equation}\label{eq-PI_score}
\noindent
PI=\frac{\#~Ki67^{+}~cells}{\#(Ki67^{+}~+~Ki67^{-})~cells}
\end{equation} To quantify nuclei detection performance, the F1-score is used:
\begin{equation}\label{eq-F1_score}
\noindent
F1=\frac{2\times TP}{2\times TP + FP + FN}
\end{equation}
where TP, FP, and FN represent the number of true positives, false positives, and false negatives, respectively.  For images that only have PI, the difference in the computed and pathologist given PI is investigated.

\section{Deep Learning Frameworks}

U-Net~\cite{ronneberger} and its derivatives such as MultiresU-Net~\cite{ibtehaz}, DenseU-Net~\cite{guan}, and piNet~\cite{geread_pinet} have shown outstanding performances over several image segmentation and detection domains. These architectures are often optimized in their initial introduced domain of applications. Consequently, when they are adopted to perform tasks in a new domain, the results may not be as promising as expected. Nuclear detection in pathology images requires architectures that concentrate on high-resolution details as fine as nuclear size and smaller. As an example, U-Net that is composed of convolutional neural networks, may not be able to sufficiently recover details when dense pathology images are studied. One reason is that successive convolutional layers and early maxpoolings result in removing high-resolution information and fine details that are important for quantifying small nuclei. 

To address this problem, we introduce an architecture referred to as UV-Net which focuses on preserving dense features. Figure~\ref{fig-Architecture}a shows the full architecture, where 3$\times$3 convolutional layers used in U-Net are replaced by V-Blocks, inspired by the efficiency of dense connections \cite{guan}. Each V-Block expands an input with n channels to output with 2n channels (creating a "V" shape) through four successive stages. Two hyperparameters, f and k, are defined for each V-Block where they are equal to the number of input channels, and the output channels at the end of each stage, respectively. Figure~\ref{fig-Architecture}b shows a V-Block wherein $f=16$ and $k=4$. In each stage, the input feature is processed by a 1$\times$1 convolution with $f=16$ filters, then transformed to the output with $k=4$ filters. The output of this step is concatenated to the input, creating a matrix with 20 filters which are fed to the second stage. This process is repeated for a total of four times to generate an output with 2$\times$f filters. The successive concatenations prevent losing features obtained from earlier layers, specifically nuclear ones. Results from the proposed UV-Net are compared to U-Net, MultiresU-Net and DenseU-Net, as well as piNet which is developed specifically for Ki67 tasks \cite{geread_pinet}.

\newcommand{\subf}[2]{%
  {\small\begin{tabular}[t]{@{}c@{}}
  #1\\#2
  \end{tabular}}%
}
\begin{figure}
\centering
\begin{tabular}{c c}

\subf{\includegraphics[width=0.6\linewidth]{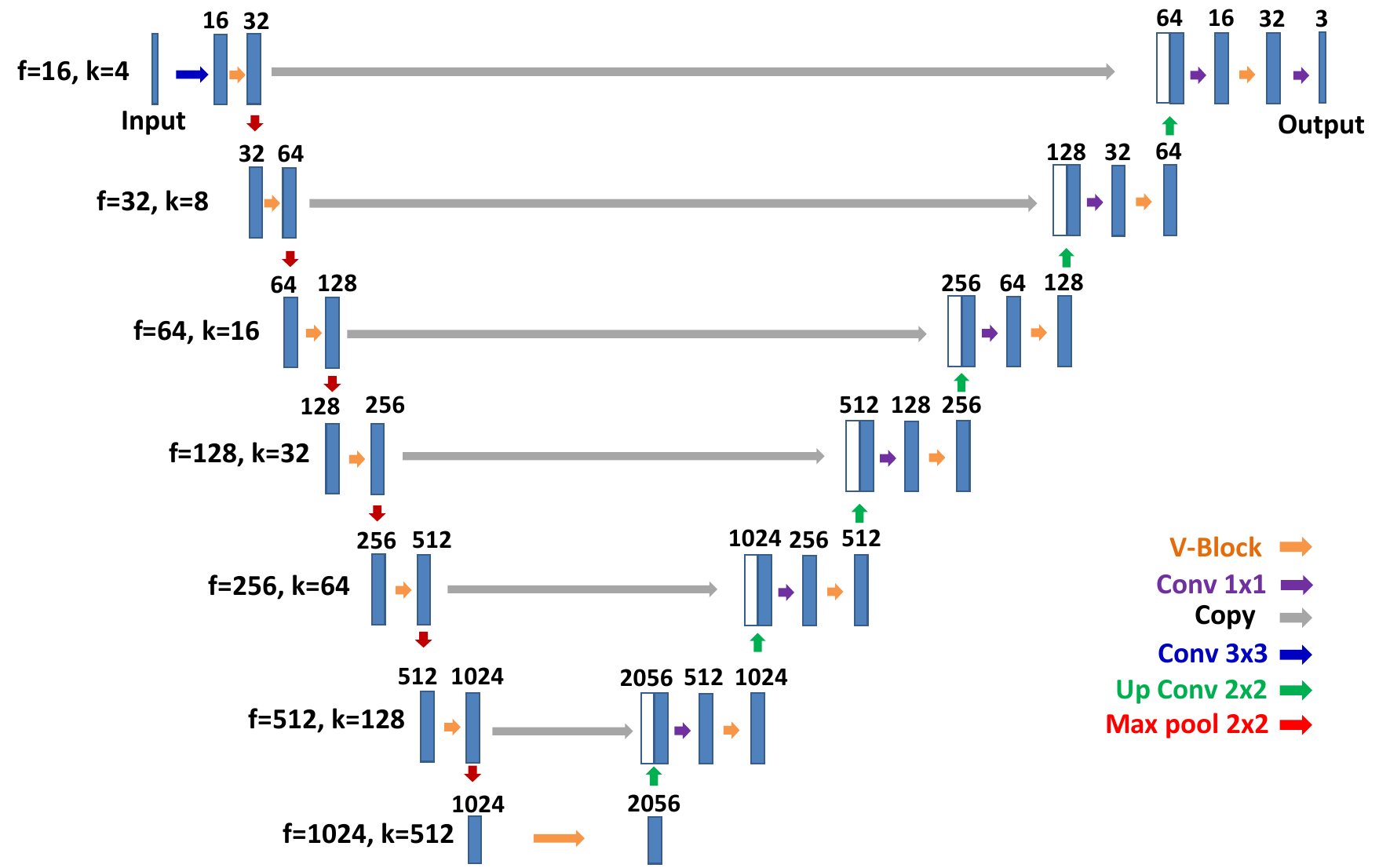}}
     {a}
&
\subf{\includegraphics[width=0.35\linewidth]{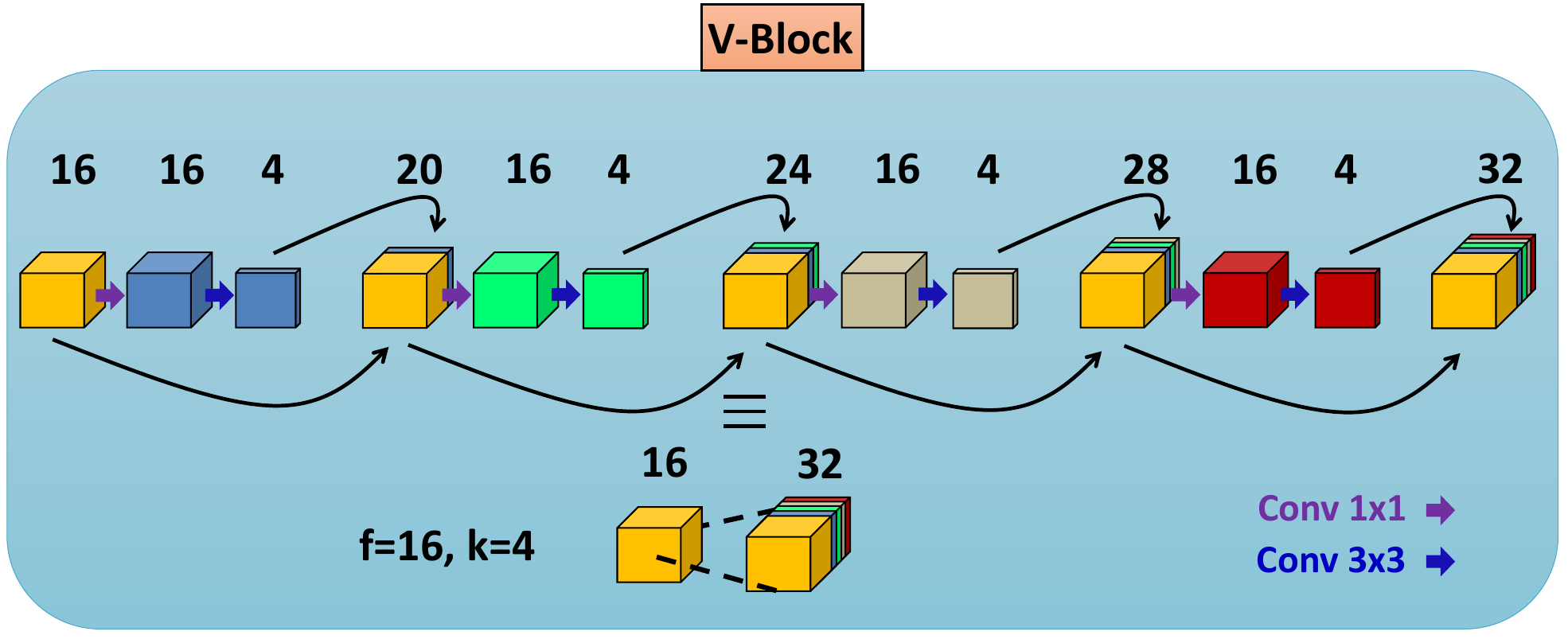}}
     {b}
\\
\end{tabular}
\caption{Architecture. a. UV-Net architecture that is made up of encoding and decoding arms. In both arms, V-Blocks are used to preserve features. The input and output images have 256$\times$256$\times$3 dimensions. b. One V-Block example where f=16, and k=4, composed of four stages. The output of each stage is concatenated with earlier outputs to avoid losing features.}
\label{fig-Architecture}
\end{figure}

\section{Results}

\begin{figure}[t]
\begin{center}
   \includegraphics[width=0.9\linewidth]{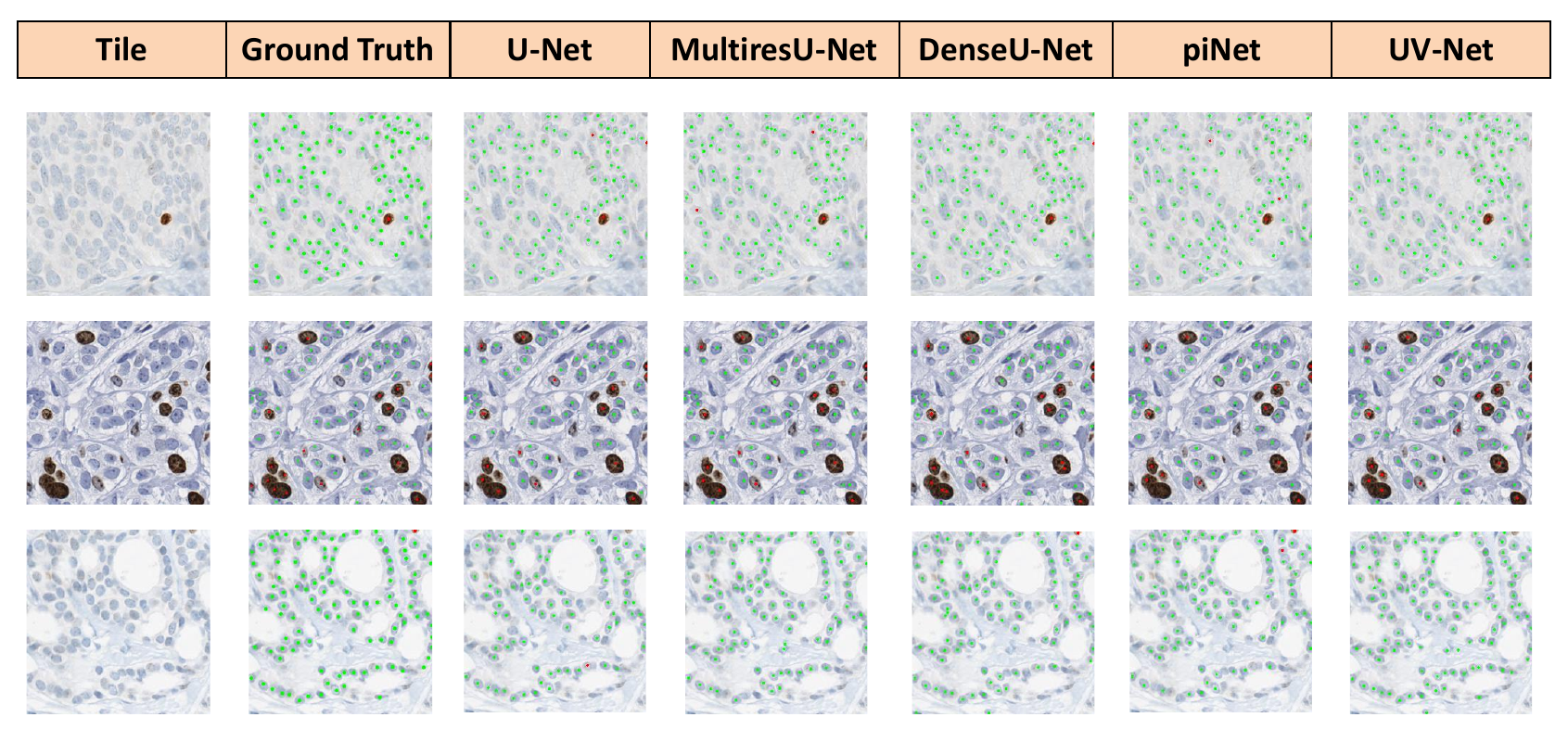}
   \caption{Prediction results. Samples of tiles belong to SMH (1$^{st}$ and 3$^{rd}$ rows) and Deepslides (2$^{nd}$ row) data, with their corresponding ground truths and predictions.}
\label{fig-Tiles}
\end{center}
\end{figure}

This section presents the results of U-Net, MultiresU-Net, DenseU-Net, piNet, and UV-Net for nuclei detection and PI estimation on five multicentre datasets. All architectures were trained using the SMH and Deepslides patches. Huber loss function was used for all architectures to regress and predict the centroid of the nuclei. Data augmentations such as horizontal and vertical flips, as well as scaling are used. All experiments were conducted on the same machine with an NVIDIA GeForce RTX 2080 Ti. A total of 100 epochs were run with an Adam optimizer, batch size=4, and learning rate=10$^{-3}$. The predictions obtained for three patches from Deepslides and SMH over all five architectures are shown in Figure~\ref{fig-Tiles} for visualization purposes.\\ 

\begin{figure}
\centering
\begin{tabular}{c c}

\subf{\includegraphics[width=0.4\linewidth]{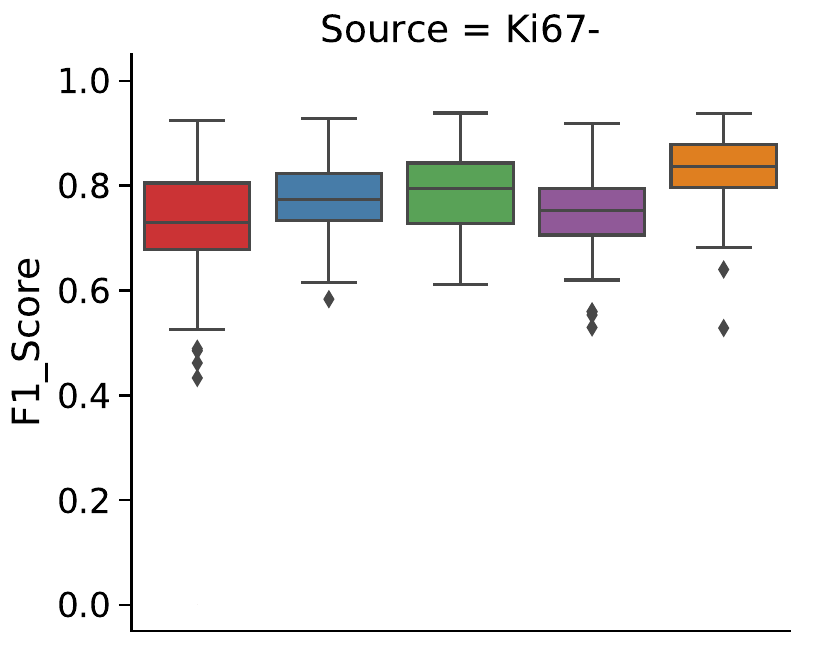}}
     {a}
&
\subf{\includegraphics[width=0.5\linewidth]{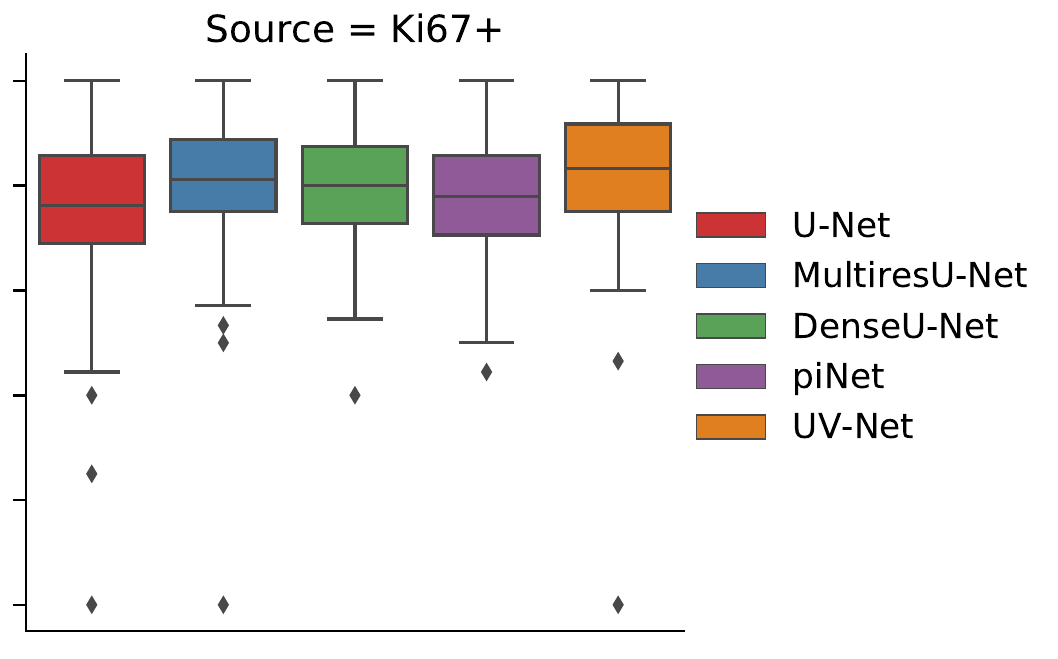}}
     {b}
\\
\end{tabular}
\caption{Test results. F1-score results on the SMH and Deepslides dataset for five architectures. a. On Ki67$^{-}$ b. On Ki67$^{+}$.}
\label{fig-F1-score_Deepslides_SMH}
\end{figure}

\indent Figure~\ref{fig-F1-score_Deepslides_SMH} shows the F1-score accuracy results of nuclei detection for both Ki67$^{-}$ and Ki67$^{+}$ cells obtained by multiple architectures when applied to Deepslides and SMH held out testing data. Table~\ref{Table-Deepslide_SMH} also shows the mean F1-score for both Ki67$^{-}$ and Ki67$^{+}$ cells over multiple architectures applied to Deepslides-SMH held out testing data. It is seen that our proposed architecture, UV-Net, outperforms other architectures by 4.69-10.76\% on Ki67$^{-}$ and 1.19-6.56\% on Ki67$^{+}$. 


%

\begin{table}
\begin{center}
\begin{tabular}{|c|c|c|}
\hline
Architecture & F1(Ki67$^{-}$) & F1(Ki67$^{+}$) \\
\hline\hline
U-Net & 0.7257 & 0.7545\\
MultiresU-Net & 0.7764 & 0.8073\\
DenseU-Net & 0.7864 & 0.8082\\
piNet & 0.7493 & 0.7830\\
UV-Net & \bf{0.8333} & \bf{0.8201}\\
\hline
\end{tabular}
\end{center}
\caption{Average of F1-score results on Deepslide SMH test data for Ki67$^{-}$ and Ki67$^{+}$.}
\label{Table-Deepslide_SMH}
\end{table}

\begin{figure}
\centering
\begin{tabular}{c c}

\subf{\includegraphics[width=0.4\linewidth]{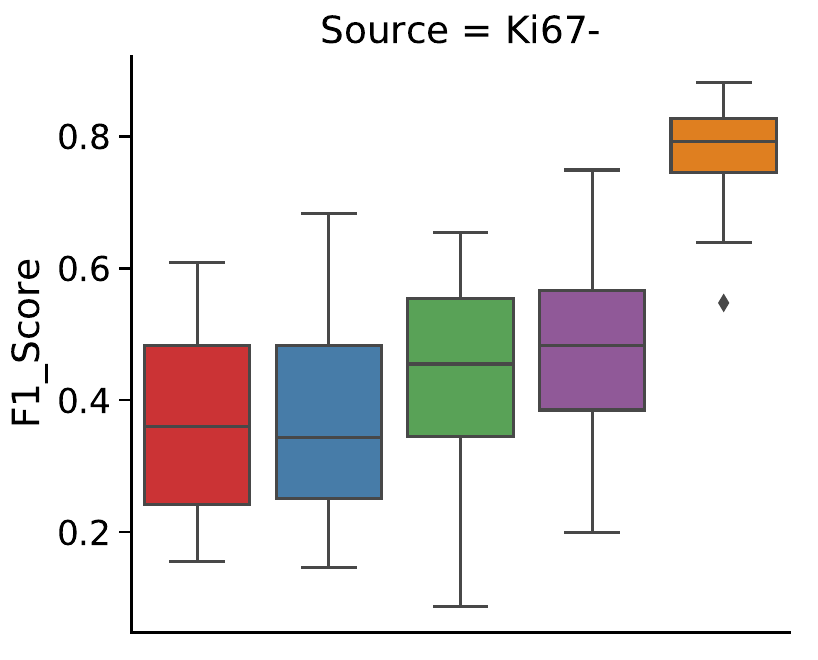}}
     {a}
&
\subf{\includegraphics[width=0.5\linewidth]{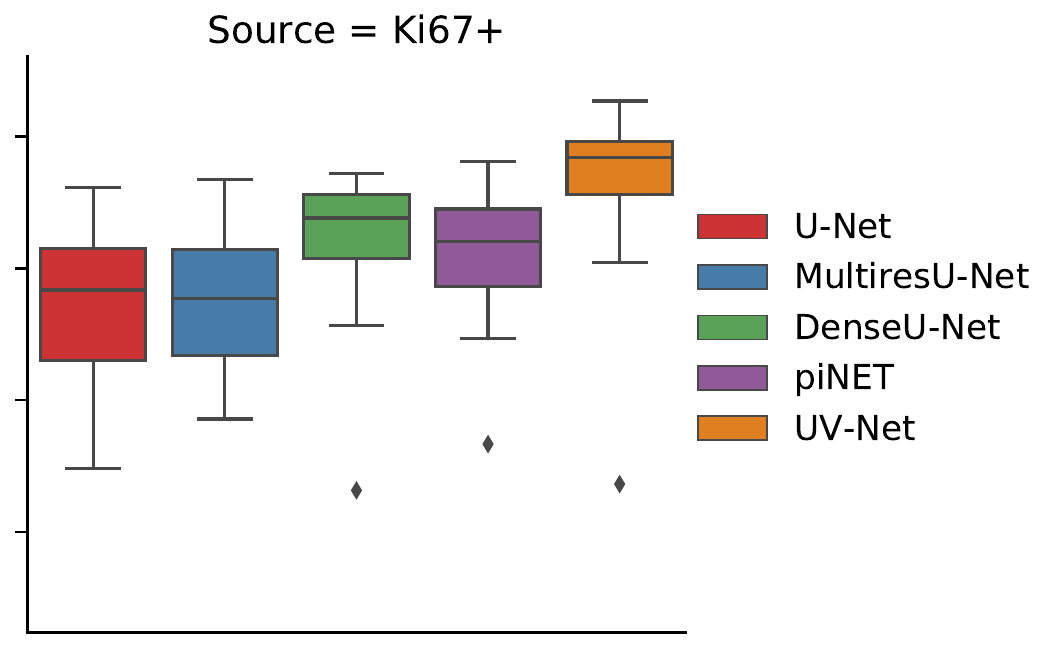}}
     {b}
\\
\end{tabular}
\caption{Test results. F1-score results on OVC TMA dataset for five architectures. a. On Ki67$^{-}$ b. On Ki67$^{+}$ cells.}
\label{fig-F1-score_OVC}
\end{figure}

\begin{figure}
\centering
\begin{tabular}{c c}

\subf{\includegraphics[width=0.45\linewidth]{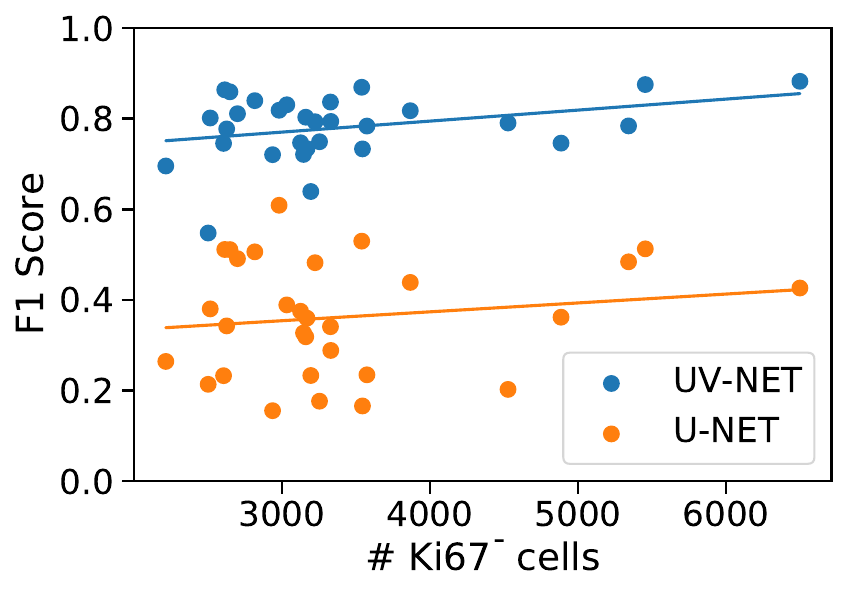}}
     {a}
&
\subf{\includegraphics[width=0.45\linewidth]{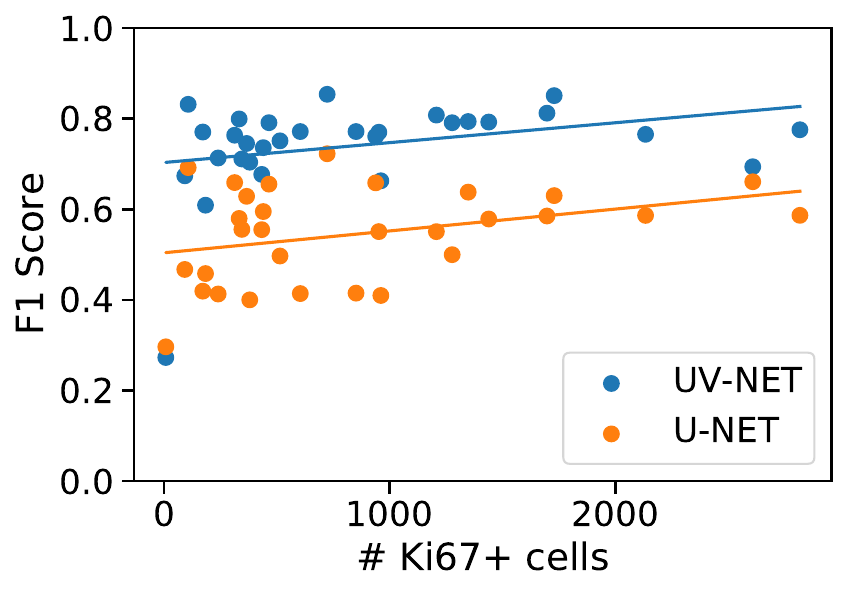}}
     {b}
\\
\end{tabular}
\caption{Prediction distribution. Distribution of predicted results of U-Net and UV-Net per each TMA image for the OVC dataset with corresponding F1-scores versus a. total number of Ki67$^{-}$ cells b. total number of Ki67$^{+}$ cells.}
\label{fig-F1_distribution}
\end{figure}

\indent To assess the generalization capabilities, the trained models are evaluated on images from multiple institutions. 
The first dataset is from Protein Atlas, which has PI ranges for ground truths, [0-25], [25-75], and [75-100]$\%$. Therefore, the computed PIs from trained networks are compared with the ground truth PI ranges to evaluate performances. Table~\ref{Table-Atlas} shows the average PI accuracy calculated by comparing whether a predicted PI fell in the range of its corresponding ground truth. UV-Net by far achieves the highest PI estimation accuracy as compared to other architectures. The second dataset is from the OVC with the F1-scores shown in Figure~\ref{fig-F1-score_OVC}. As observed, UV-Net significantly outperforms other architectures with an average F1-score higher than other networks by 31 - 42$\%$ on Ki67$^{-}$ and 9 - 20$\%$ on Ki67$^{+}$ cells. The F1-score distribution versus the number of Ki67$^{-}$ and Ki67$^{+}$ cells per each OVC TMA image is plotted in Figure~\ref{fig-F1_distribution}. The distribution shows the results for two architectures, U-Net and UV-Net, where the size of all images are ~1400$\times$1400. It is seen that for nearly all images, UV-Net outperforms U-Net. More importantly, UV-Net's performance does not degrade with an increasing number of cells. The F1 performance across the Ki67$^{+}$ and Ki67$^{-}$ cells is more consistent with UV-Net as compared to U-Net which has a significantly lower average F1-score in the dense Ki67$^{-}$ cells. This is due to the fact that in a single TMA the populations of Ki67$^{-}$ cells, on average, is higher than that of Ki67$^{+}$ cells. In contrast, UV-Net is capable of high performance in images with high amounts of Ki67$^{-}$ cells, indicating that in densely populated regions with clusters of cells, noticeable performance improvement is observed. \\

\begin{table}
\begin{center}
\begin{tabular}{|c|c|}
\hline
Architecture & PI Accuracy \\
\hline\hline
U-Net & 0.5721\\
MultiresU-Net & 0.4135\\
DenseU-Net & 0.4576\\
piNet & 0.6927\\
UV-Net & \bf{0.7631}\\
\hline
\end{tabular}
\end{center}
\caption{Average proliferation index results on protein Atlas TMA dataset.}
\label{Table-Atlas}
\end{table}

\begin{figure}
\centering
\begin{tabular}{c c}

\subf{\includegraphics[width=0.4\linewidth]{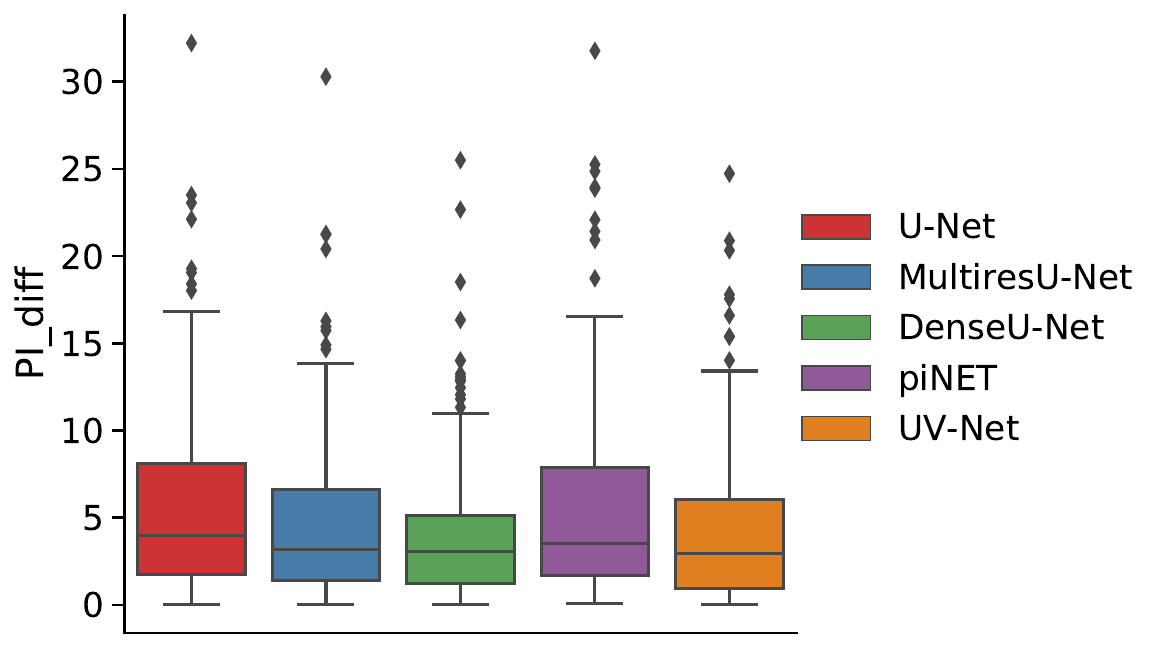}}
     {a}
&
\subf{\includegraphics[width=0.4\linewidth]{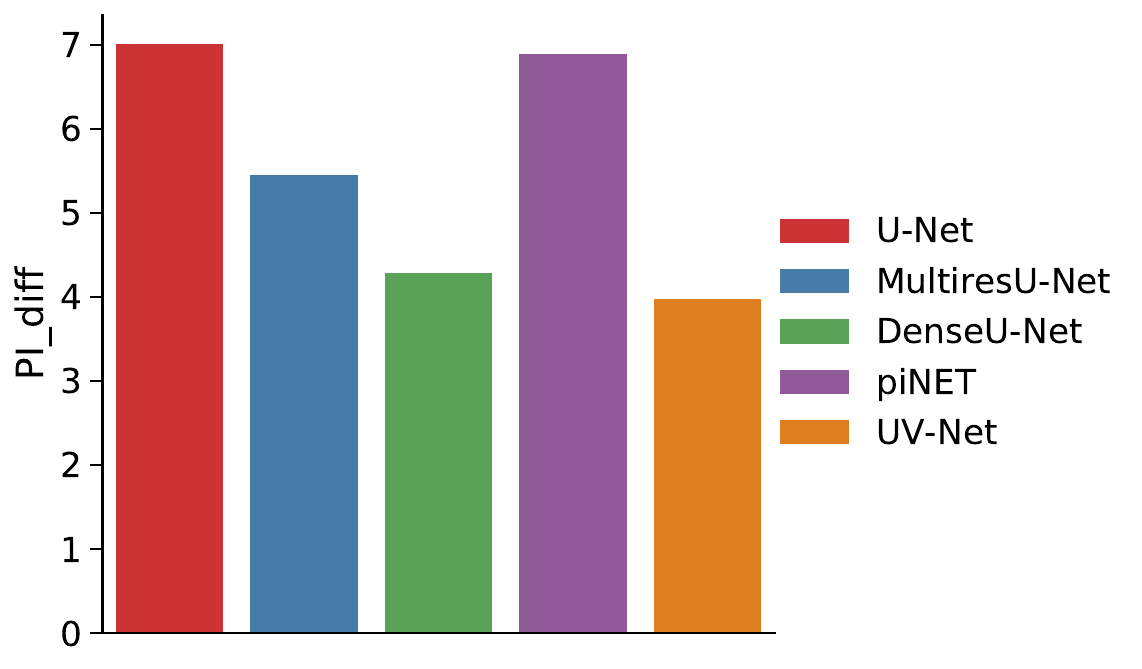}}
     {b}
\\
\end{tabular}
\caption{PI difference for five architectures on UHN TMAs. a. PI difference distributions, b. Average PI differences.}
\label{fig-F1-score_UHN}
\end{figure}

\indent The third dataset is from UHN. Our pathologist scored TMA cores based on counting 200 cells in tumour hot spots, where available, then an average PI across all three cores are taken for each patient. This was done to account for tumour heterogeneity. The trained networks, however, scan the entire TMA regions to obtain comprehensive detection. Similar to the pathologist's procedure, an averaging is done over all TMAs' proliferation indexes belong to a patient to obtain a single value. The results are then compared with provided ground truths. Figure~\ref{fig-F1-score_UHN} shows the PI differences between ground truths and predicted PIs per patient. It is seen that all architectures report an average PI difference value of less than 7$\%$. UV-Net and DenseU-Net achieve the best average PIs equal to 3.98$\%$ and 4.28$\%$, respectively. DenseU-Net shows a lower variance over all data. Hencem, the coefficient of variations (CoV) are calculated to account for both PI mean and standard deviation values. UV-Net scores CoV=1.08 and DenseU-Net scores CoV=1.12, showing that UV-Net performs slightly higher than DenseU-Net on this dataset as well.

\begin{figure}
\centering
\begin{tabular}{c c}

\subf{\includegraphics[width=0.4\linewidth]{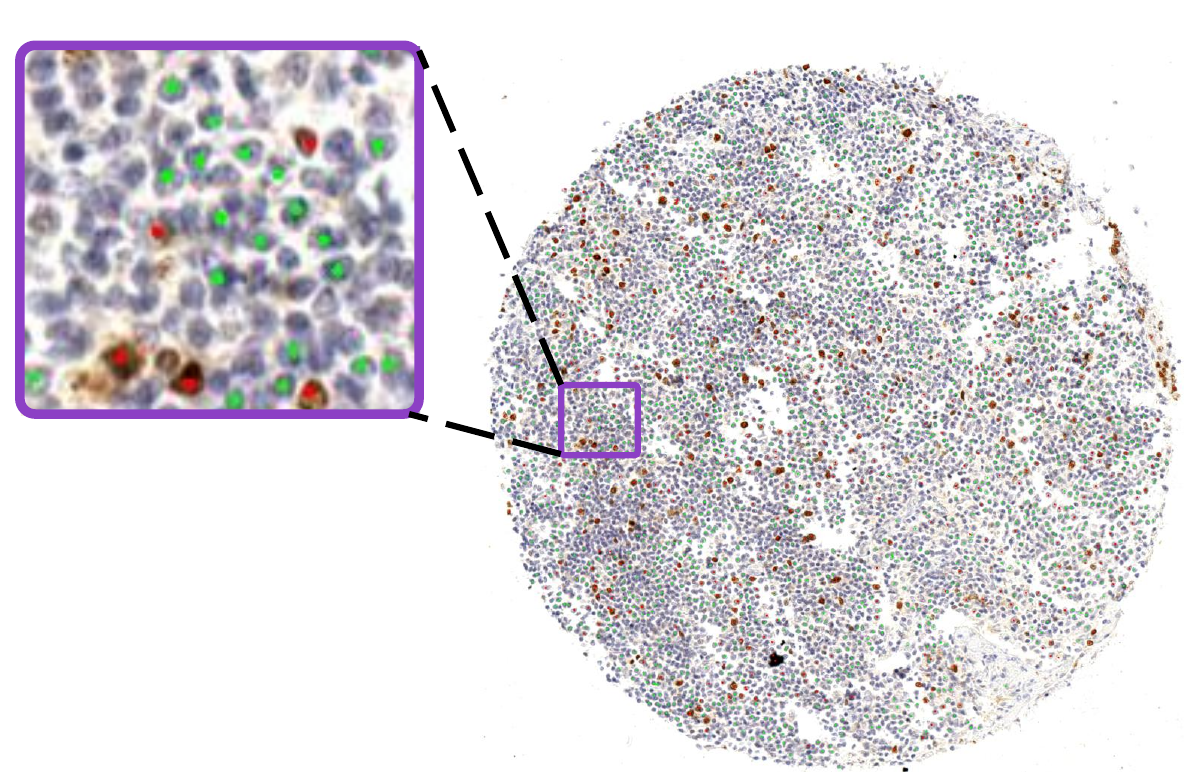}}
     {a}
&
\subf{\includegraphics[width=0.4\linewidth]{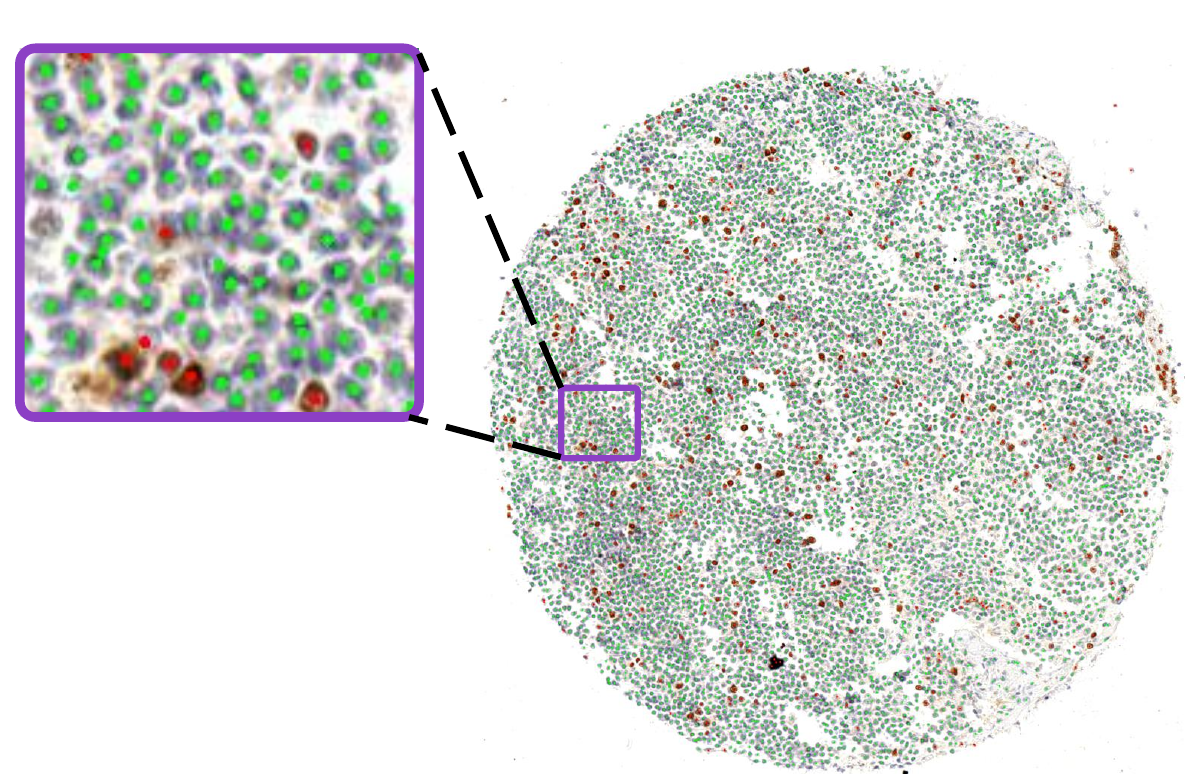}}
     {b}
\\
\end{tabular}
\caption{Qualitative predicted results on the OVC dataset. One TMA image, predicted by a. U-Net b. UV-Net.}
\label{fig-TMA_OVC}
\end{figure}

\begin{figure}
\centering
\begin{tabular}{c c}

\subf{\includegraphics[width=0.4\linewidth]{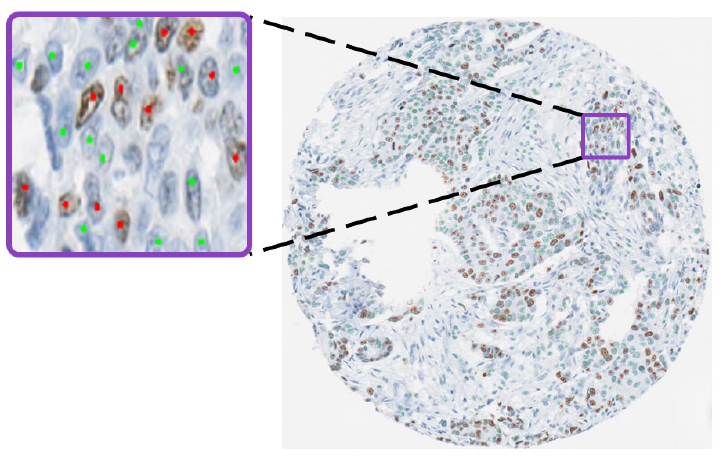}}
     {a}
&
\subf{\includegraphics[width=0.4\linewidth]{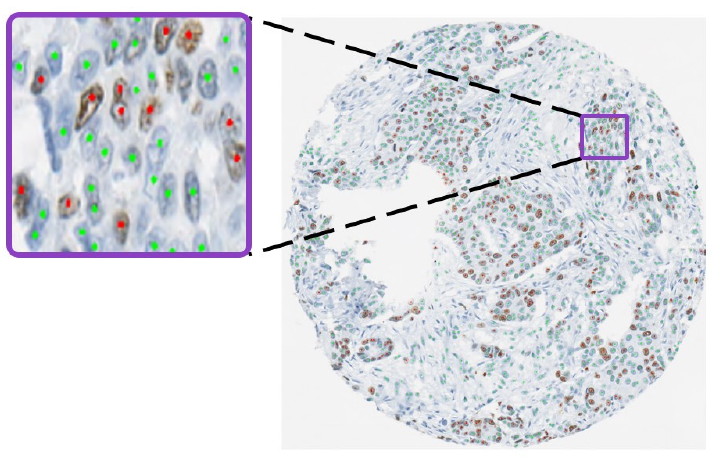}}
     {b}
\\
\end{tabular}
\caption{Qualitative predicted results on the UHN dataset. One TMA image, predicted by a. U-Net b. UV-Net.}
\label{fig-TMA_UHN}
\end{figure}

\indent To qualitatively compare the performances of architectures, specifically on regions with clustered nuclei, one OVC TMA image of size 1400$\times$1400 with 6498 Ki67$^{-}$ and 514 Ki67$^{+}$ cells is selected. Figure~\ref{fig-TMA_OVC} shows the predicted TMA image obtained from U-Net and UV-Net. It is seen in  Figure~\ref{fig-TMA_OVC}a that U-Net sparsely detected nuclei and a large portion of them are missed. In contrast, UV-Net successfully showed a higher performance in this dense medium as, Figure~\ref{fig-TMA_OVC}b. Additionally, U-Net shows an inconsistent performance in detecting nuclei, whereas UV-Net reports more consistent results. The acquired F1-scores obtained by U-Net in this case for Ki67$^{-}$ and Ki67$^{+}$ cells are 0.4261 and 0.4968, respectively. UV-Net shows a much higher performance by reporting 0.8824 for Ki67$^{-}$ and 0.7510 for Ki67$^{+}$ cells. Figure~\ref{fig-TMA_UHN} shows the qualitative comparisons between U-Net and UV-Net when a TMA image from the UHN dataset is considered. The image has a size of 2000$\times$2000 where the average number of Ki67$^{-}$ and Ki67$^{+}$ cells obtained by all architectures are 743 and 321, respectively. Similar conclusions to that of Figure~\ref{fig-TMA_OVC} remains valid in this case.\\

\section{Conclusion}

This paper focused on evaluating multiple widely-used architectures for the purpose of nuclear detection on breast cancer histology images with dense medium. A new architecture, referred to as UV-Net, was introduced in this paper specifically to restore nuclear features in dense medium by preserving high-resolution details. Full comparisons with widely used networks were presented. UV-Net shows an outstanding performance in preserving and detecting nuclei in dense medium, as compared to other architectures. The generalizability degree of various architectures is evaluated by testing architectures on multi-institutional datasets gathered from five different institutes. The dataset ranges from WSI patches to tissue micro-arrays from human and canine mammary breast cancer images. UV-Net showed consistently higher performance on all datasets.

\bibliography{UV_NET_Nuclei} 
\bibliographystyle{spiebib} 
\end{document}